\documentclass[aps,prl,showpacs,twocolumn,superscriptaddress]{revtex4}
\usepackage{graphicx}
\usepackage{dcolumn}
\usepackage{bm}
\usepackage{amsmath, amssymb}%

\newcommand{\eepp}{\mbox{$(e,e'pp)$}}
\newcommand{\eep}{\mbox{($e,e'p$)}} 

\newcommand{\eeppn}{\mbox{$(e,e'pp)n$}}
 
\newcommand{\Het}{$^3$He}

\begin{document}


\title{Tensor Correlations Measured in $^3$He$(e,e'pp)n$}

\newcommand*{\ANL}{Argonne National Laboratory, Argonne, Illinois 60441}
\newcommand*{\ANLindex}{1}
\affiliation{\ANL}
\newcommand*{\ASU}{Arizona State University, Tempe, Arizona 85287-1504}
\newcommand*{\ASUindex}{2}
\affiliation{\ASU}
\newcommand*{\CSUDH}{California State University, Dominguez Hills, Carson, CA 90747}
\newcommand*{\CSUDHindex}{3}
\affiliation{\CSUDH}
\newcommand*{\CMU}{Carnegie Mellon University, Pittsburgh, Pennsylvania 15213}
\newcommand*{\CMUindex}{4}
\affiliation{\CMU}
\newcommand*{\CUA}{Catholic University of America, Washington, D.C. 20064}
\newcommand*{\CUAindex}{5}
\affiliation{\CUA}
\newcommand*{\SACLAY}{CEA, Centre de Saclay, Irfu/Service de Physique Nucl\'eaire, 91191 Gif-sur-Yvette, France}
\newcommand*{\SACLAYindex}{6}
\affiliation{\SACLAY}
\newcommand*{\CNU}{Christopher Newport University, Newport News, Virginia 23606}
\newcommand*{\CNUindex}{7}
\affiliation{\CNU}
\newcommand*{\UCONN}{University of Connecticut, Storrs, Connecticut 06269}
\newcommand*{\UCONNindex}{8}
\affiliation{\UCONN}
\newcommand*{\EDINBURGH}{Edinburgh University, Edinburgh EH9 3JZ, United Kingdom}
\newcommand*{\EDINBURGHindex}{9}
\affiliation{\EDINBURGH}
\newcommand*{\FU}{Fairfield University, Fairfield CT 06824}
\newcommand*{\FUindex}{10}
\affiliation{\FU}
\newcommand*{\FIU}{Florida International University, Miami, Florida 33199}
\newcommand*{\FIUindex}{11}
\affiliation{\FIU}
\newcommand*{\FSU}{Florida State University, Tallahassee, Florida 32306}
\newcommand*{\FSUindex}{12}
\affiliation{\FSU}
\newcommand*{\GWUI}{The George Washington University, Washington, DC 20052}
\newcommand*{\GWUIindex}{13}
\affiliation{\GWUI}
\newcommand*{\ISU}{Idaho State University, Pocatello, Idaho 83209}
\newcommand*{\ISUindex}{14}
\affiliation{\ISU}
\newcommand*{\INFNFR}{INFN, Laboratori Nazionali di Frascati, 00044 Frascati, Italy}
\newcommand*{\INFNFRindex}{15}
\affiliation{\INFNFR}
\newcommand*{\INFNGE}{INFN, Sezione di Genova, 16146 Genova, Italy}
\newcommand*{\INFNGEindex}{16}
\affiliation{\INFNGE}
\newcommand*{\INFNRO}{INFN, Sezione di Roma Tor Vergata, 00133 Rome, Italy}
\newcommand*{\INFNROindex}{17}
\affiliation{\INFNRO}
\newcommand*{\ORSAY}{Institut de Physique Nucl\'eaire ORSAY, Orsay, France}
\newcommand*{\ORSAYindex}{18}
\affiliation{\ORSAY}
\newcommand*{\ITEP}{Institute of Theoretical and Experimental Physics, Moscow, 117259, Russia}
\newcommand*{\ITEPindex}{19}
\affiliation{\ITEP}
\newcommand*{\JMU}{James Madison University, Harrisonburg, Virginia 22807}
\newcommand*{\JMUindex}{20}
\affiliation{\JMU}
\newcommand*{\KNU}{Kyungpook National University, Daegu 702-701, Republic of Korea}
\newcommand*{\KNUindex}{21}
\affiliation{\KNU}
\newcommand*{\LPSC}{LPSC, Universite Joseph Fourier, CNRS/IN2P3, INPG, Grenoble, France}
\newcommand*{\LPSCindex}{22}
\affiliation{\LPSC}
\newcommand*{\UNH}{University of New Hampshire, Durham, New Hampshire 03824-3568}
\newcommand*{\UNHindex}{23}
\affiliation{\UNH}
\newcommand*{\NSU}{Norfolk State University, Norfolk, Virginia 23504}
\newcommand*{\NSUindex}{24}
\affiliation{\NSU}
\newcommand*{\OHIOU}{Ohio University, Athens, Ohio  45701}
\newcommand*{\OHIOUindex}{25}
\affiliation{\OHIOU}
\newcommand*{\ODU}{Old Dominion University, Norfolk, Virginia 23529}
\newcommand*{\ODUindex}{26}
\affiliation{\ODU}
\newcommand*{\RPI}{Rensselaer Polytechnic Institute, Troy, New York 12180-3590}
\newcommand*{\RPIindex}{27}
\affiliation{\RPI}
\newcommand*{\URICH}{University of Richmond, Richmond, Virginia 23173}
\newcommand*{\URICHindex}{28}
\affiliation{\URICH}
\newcommand*{\ROMAII}{Universita' di Roma Tor Vergata, 00133 Rome Italy}
\newcommand*{\ROMAIIindex}{29}
\affiliation{\ROMAII}
\newcommand*{\MSU}{Skobeltsyn Nuclear Physics Institute, Skobeltsyn Nuclear Physics Institute, 119899 Moscow, Russia}
\newcommand*{\MSUindex}{30}
\affiliation{\MSU}
\newcommand*{\SCAROLINA}{University of South Carolina, Columbia, South Carolina 29208}
\newcommand*{\SCAROLINAindex}{31}
\affiliation{\SCAROLINA}
\newcommand*{\JLAB}{Thomas Jefferson National Accelerator Facility, Newport News, Virginia 23606}
\newcommand*{\JLABindex}{32}
\affiliation{\JLAB}
\newcommand*{\UNIONC}{Union College, Schenectady, NY 12308}
\newcommand*{\UNIONCindex}{33}
\affiliation{\UNIONC}
\newcommand*{\UTFSM}{Universidad T\'{e}cnica Federico Santa Mar\'{i}a, Casilla 110-V Valpara\'{i}so, Chile}
\newcommand*{\UTFSMindex}{34}
\affiliation{\UTFSM}
\newcommand*{\GLASGOW}{University of Glasgow, Glasgow G12 8QQ, United Kingdom}
\newcommand*{\GLASGOWindex}{35}
\affiliation{\GLASGOW}
\newcommand*{\VIRGINIA}{University of Virginia, Charlottesville, Virginia 22901}
\newcommand*{\VIRGINIAindex}{36}
\affiliation{\VIRGINIA}
\newcommand*{\WM}{College of William and Mary, Williamsburg, Virginia 23187-8795}
\newcommand*{\WMindex}{37}
\affiliation{\WM}
\newcommand*{\YEREVAN}{Yerevan Physics Institute, 375036 Yerevan, Armenia}
\newcommand*{\YEREVANindex}{38}
\affiliation{\YEREVAN}
\newcommand*{\CANISIUS}{Canisius College, Buffalo, NY 14208}
\newcommand*{\CANISIUSindex}{39}
\affiliation{\CANISIUS}

\newcommand*{\NOWGWUII}{The George Washington University2, Washington, DC 20052}
\newcommand*{\NOWEDINBURGH}{Edinburgh University, Edinburgh EH9 3JZ, United Kingdom}
\newcommand*{\NOWVIRGINIA}{University of Virginia, Charlottesville, Virginia 22901}

  
\author{H. Baghdasaryan}
     \altaffiliation[Current address: ]{\NOWVIRGINIA}
     \affiliation{\ODU}
\author{L.B.~Weinstein}
\email[Contact Author \ ]{weinstein@odu.edu}
     \affiliation{\ODU}
\author{J.M.~Laget}
\affiliation{\JLAB}
\author {K.P. ~Adhikari} 
\affiliation{\ODU}
\author {M.~Aghasyan} 
\affiliation{\INFNFR}
\author {M.~Amarian} 
\affiliation{\ODU}
\author {M.~Anghinolfi} 
\affiliation{\INFNGE}
\author {H.~Avakian} 
\affiliation{\JLAB}
\affiliation{\INFNFR}
\author {J.~Ball} 
\affiliation{\SACLAY}
\author {M.~Battaglieri} 
\affiliation{\INFNGE}
% ...

\collaboration{The CLAS Collaboration}
     \noaffiliation

\date{\today}

\begin{abstract}
We have measured the $^3$He\mbox{$(e,e'pp)n$} reaction at an incident energy of 4.7 GeV over
a wide kinematic range.  We identified spectator correlated $pp$ and $pn$ nucleon pairs using
kinematic cuts and measured their relative and total momentum distributions.  This is the first measurement of the ratio of $pp$ to $pn$
pairs as a function of pair total momentum, $p_{tot}$.  For pair relative momenta between 0.3 and 0.5 GeV/c, the ratio is
very small at low $p_{tot}$ and rises to approximately 0.5 at large
$p_{tot}$.  This shows the dominance of tensor over central correlations at this
relative momentum.
\end{abstract}


\maketitle


In order to understand the structure of the nucleus, we need to
understand both the independent motion of individual nucleons and
the corrections to that simple picture.  Single nucleon momentum
distributions have been measured in electron-proton knockout
reactions, $(e,e'p)$, and are reasonably well understood
\cite{frullani84, kelly96, gao00}.  However, only about 70\% of the
naively expected number of protons are seen.  The missing 30\% are
presumably due to nucleons in short range and long range correlations.

These nucleon-nucleon ($NN$) correlations are the next important
ingredient.  A $^{12}$C$(p,ppn)$ experiment \cite{tang03} found that low momentum
neutrons, $p_n<0.22$ GeV/c,  were emitted isotropically but that high
momentum neutrons were emitted opposite to the struck proton's missing
momentum, $\vec p_{miss}$, and were therefore the correlated
partner of the struck protons.

 
Measurements of the cross section ratios of inclusive electron scattering from
nuclei relative to deuterium, $\sigma[A(e,e')]/\sigma[d(e,e')]$,
together with calculations of deuterium show that the momentum
distributions for $p > 0.25$ GeV/c have the same shape for all nuclei and
that nucleons have between a 5\% and a 25\% 
probability of being part of a correlated pair
\cite{egiyan02,egiyan06,antonov88,forest96}.  

Thus, when a nucleon has
low momentum, $p<p_{fermi}$, its momentum is balanced by the rest of
the nucleus; however, when $p>p_{fermi}$,
its momentum is almost always balanced by only one other nucleon and the two
nucleons form a correlated pair.  These correlated pairs are
responsible for the high momentum parts of the nuclear wave function
\cite{antonov88}.  Note that these correlations can be caused by
either the central ($L=0$) or the tensor ($L=2$) parts of the $NN$ force.

Nucleons in nuclei overlap each other a significant
fraction of the time. These high momentum correlated pairs should be
at significantly higher local density than the nuclear average.  Thus,
understanding correlated $NN$ pairs will improve our
understanding of cold dense nuclear matter, neutron stars
\cite{frankfurt08}, and the EMC effect \cite{sargsian02}.

Recent measurements of direct two nucleon knockout from carbon using
protons \cite{piasetzky06} and electrons \cite{shneor07,subedi08} have
shown that the removal of a proton from the nucleus with 
$0.275 < p_{miss} < 0.550$ GeV/c is almost always accompanied by the emission
of a correlated nucleon that carries momentum roughly equal and
opposite to $\vec p_{miss}$ and that this nucleon is
almost always a neutron.  
Quantitative interpretations are complicated by
the presence of other effects, including Final State Interactions (FSI) and
two-body currents such as meson exchange currents (MEC), which add coherently to
the correlations signal \cite{janssen00}. 

A recent measurement of $^3$He$(e,e'pp)n$ \cite{niyazov03} isolated
the $NN$ correlated pairs by knocking out the third nucleon and
observing the momenta of the spectator nucleons.  It measured the $pp$
and $pn$ relative and total momentum distributions. 
Because the virtual photon was absorbed on the third nucleon, the
correlated pairs were spectators and thus the effects of two-body
currents were negligible.  However, the continuum interaction of the
spectator pair significantly reduced the cross sections and therefore
complicated the theoretical calculations \cite{glockle01,cda02,laget87}.
Thus, this type of measurement 
complements the direct knockout measurements.

This paper reports new \Het\eeppn{} results  at higher energy and
higher momentum transfer that provide a cleaner
measurement of two-nucleon momentum distributions over a wide range of
correlated pair total and relative momenta.


We measured \Het\eeppn{} at Jefferson Lab in 2002 using a 100\% duty factor, 5--10 nA beam of
4.7 GeV electrons incident on a 5-cm liquid $^3$He target.  We
detected the outgoing charged particles in the
CEBAF Large Acceptance Spectrometer (CLAS)
\cite{clas}.  

CLAS uses a toroidal magnetic field 
and six sets of
drift chambers, time-of-flight scintillation counters and
electro-magnetic calorimeters (EC) for particle identification
and trajectory reconstruction.  The polar angular acceptance is $8^o < \theta <
140^o$ and the azimuthal acceptance is 50\% at smaller polar angles, 
increasing to 80\%  at larger polar angles.  The EC was used for the
electron trigger with a threshold of $\approx 0.9$ GeV.  Regions
of non-uniform detector response were excluded by software cuts, while
acceptance and tracking efficiencies were estimated using GSIM, the
CLAS GEANT Monte-Carlo simulation.  Protons were detected down to $p_p\ge0.25$
GeV/c.  H\eep{} was measured and compared
to the world's data \cite{arrington03} to determine our electron and proton
detection efficiencies \cite{HBPhD}.

\begin{figure}[htbp]
    \includegraphics[height=2.3in]{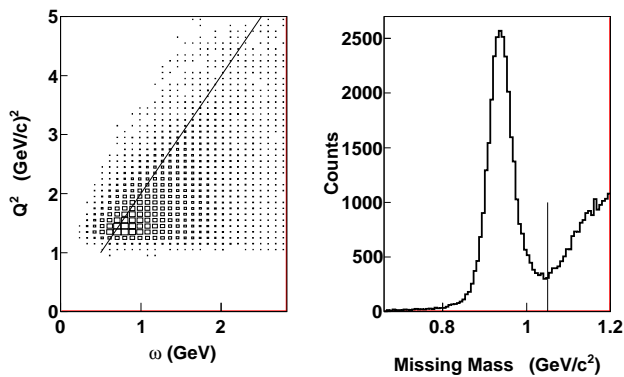} \caption{a) $Q^2$ vs.
    $\omega$ for \Het\eeppn{} events.  The line shows the quasielastic
    condition $\omega = Q^2/2m_p$.  Note the large acceptance.
    b) Missing mass for \Het\eepp$X$.  The vertical line
    indicates the neutron missing mass cut. 
    \label{fig:qomega} }
\end{figure}

We identified electrons using the energy deposited in the EC, and
protons using time-of-flight.  We identified the neutron using missing
mass to select \Het\eeppn{} events.  We eliminated target wall
interactions by selecting particles originating in the
central 4-cm of the target.  Fig.~\ref{fig:qomega} shows the electron
kinematics ($Q^2 
= \vec q\thinspace^2 - \omega^2$,
$\omega$ is the energy transfer, and $\vec q$ is the three-momentum
transfer) and missing mass distribution.  
For \Het\eeppn{} events, the momentum transfer
$Q^2$ peaks at around 1.5 (GeV/c)$^2$.  $\omega$ is
concentrated slightly above but close to quasielastic kinematics
($\omega = Q^2 / 2m_p$).

\begin{figure}[htbp]
  \includegraphics[height=1.79in,width=3.5in]{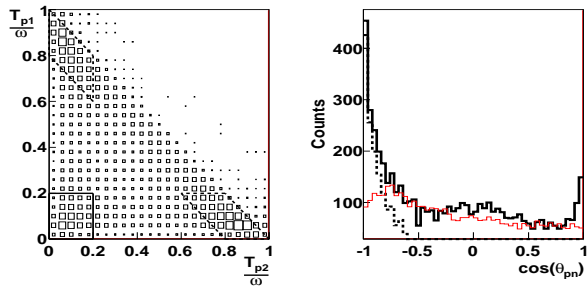} 
  \caption{a) \Het\eeppn{} lab frame ``Dalitz plot.''  $T_{p1}/\omega$
    vs. $T_{p2}/\omega$ for events with $p_N > 0.25$ GeV/c.  The solid lines indicate the
`leading $n$ plus  $pp$ pair' and the dashed lines indicate the
`leading $p$ plus  $pn$ pair' selection cuts. b) The
    cosine of the $pn$ lab frame opening angle for events with a
    leading $p$ and a $pn$ pair.  The
    thick solid line shows the uncut data, the dashed line shows the
    data cut on $p^\perp_{leading} < 0.3$ GeV/c, and the thin solid
    line (color online) shows the uncut three-body absorption
    simulation (with arbitrary normalization).  \label{fig:tkin47} }
\end{figure}
  
To understand the energy sharing in the reaction, we plotted
the lab frame kinetic energy of the first proton divided by the energy transfer
($T_{p1}/\omega$) versus that of the second proton ($T_{p2}/\omega$)
for events with nucleon momenta $p_p$ and $p_n > 0.25$ GeV/c (see Fig.{}  \ref{fig:tkin47}a).  (The assignment of protons
1 and 2 is arbitrary.  Events with $T_{p1}/\omega + T_{p2}/\omega > 1$ are
non-physical and are due to the experimental resolution.)  
There are three peaks at the three corners of the plot, corresponding
to events where two nucleons each have less than 20\% of $\omega$ and
the third `leading' nucleon has the remainder.  We selected these peaks,
as shown in Fig.{}  \ref{fig:tkin47}a.

Fig.{} \ref{fig:tkin47}b shows the opening angle for $pn$ pairs with a
leading proton (the $pp$ pair opening angle is almost identical).
Note the large peak at 180$^o$.  The peak is
not due to the cuts, since we do not see it in a simulation of
three-body absorption of the virtual photon followed by phase space
decay \cite{pdg-phasespace}. It is also not due to the CLAS acceptance
since we see it for both $pp$ and $pn$ pairs.  This back-to-back peak
is a very strong indication of correlated $NN$ pairs.

Now that we have identified correlated pairs, we want to study them.
To reduce the effects of final state rescattering, we required
the perpendicular momentum (relative to $\vec q$\thinspace) of the
leading nucleon, $p^\perp_{leading} < 0.3$ GeV/c.  
The resulting $NN$ pair opening angle distribution is almost entirely
back-to-back (see Fig.{} \ref{fig:tkin47}b).  The neutron of the $pn$
pair is distributed almost isotropically with respect to $\vec q$.
The pair average total momentum parallel to $\vec q$
($\sim 0.1$ GeV/c) is also much smaller than the average momentum
transfer ($\sim 1.6$ GeV/c).  These
show that the $NN$ pairs are predominantly spectators and
that their measured momentum distribution reflects their initial
momentum distribution.

\begin{figure}[htbp]
   \includegraphics[width=3.5in,height=3.5in]{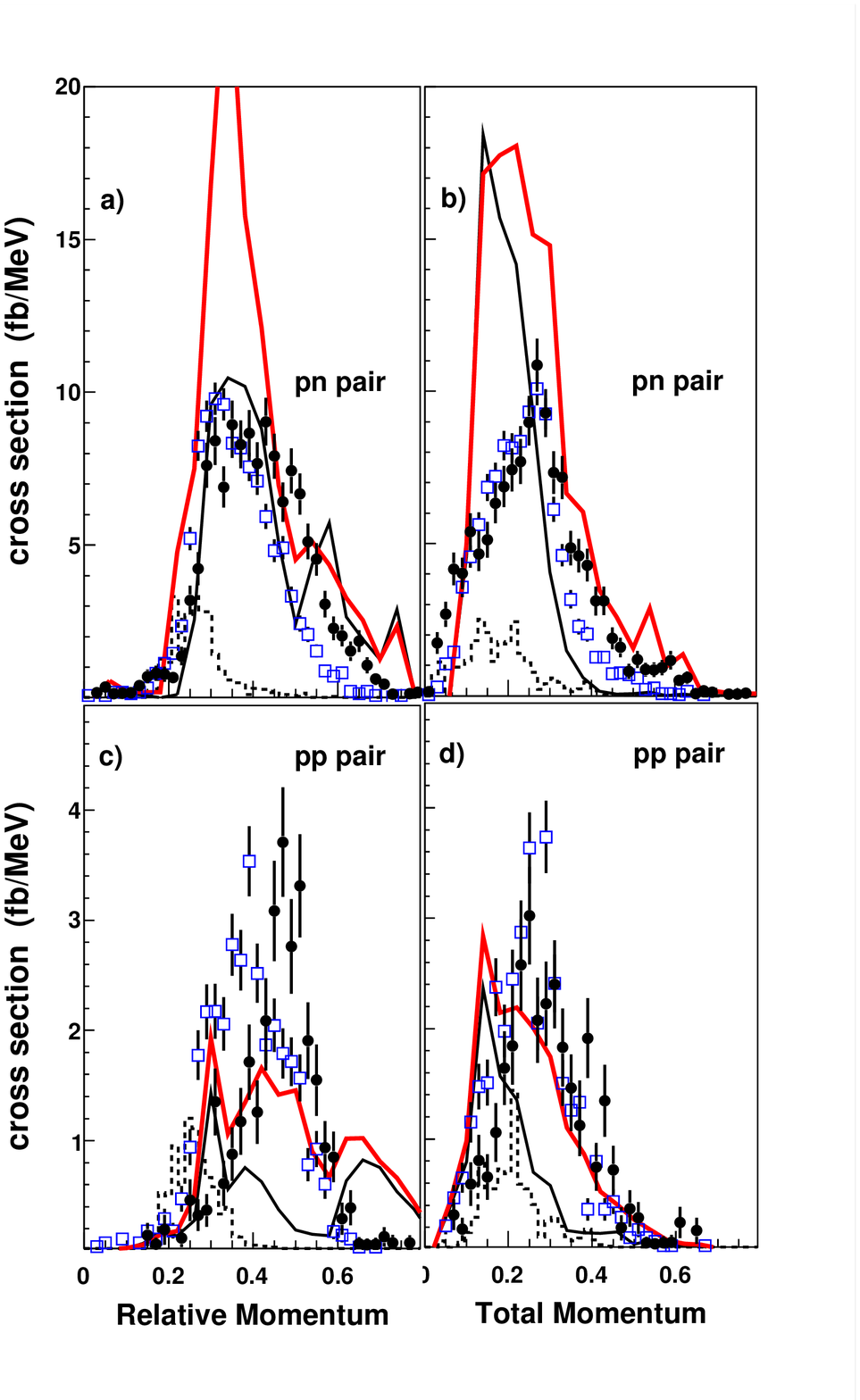}
    \caption{a) Cross section vs. $pn$ pair $p_{rel}$.  Solid points show these data ($Q^2 \sim 1.5$ GeV$^2$), open squares (blue online)
    show  $Q^2 \sim 0.7$ GeV$^2$ data  \cite{niyazov03}, dashed histogram shows the
   Golak one-body calculation \cite{golak95}, thin solid line shows the Laget one-body
   calculation and the thick solid line (red online) shows
   the Laget  full calculation \cite{laget87,laget88,laget04}; b) the same for  $p_{tot}$; c) and d)
    the same for  $pp$ pairs. All quantities are in the lab
    frame. The $Q^2\sim 0.7$ GeV$^2$ data have been
    reduced by a factor of 5.3 (the ratio of the cross
    sections) for comparison. \label{fig:prelptot} }
\end{figure}

The resulting lab frame  relative $\vec p_{rel} = (\vec p_1 - \vec p_2)/2$ and
total $\vec p_{tot} = \vec p_1 + \vec p_2$ momenta of
the  $NN$ pairs are shown in Fig.{}  \ref{fig:prelptot}.
The cross sections are integrated over the
experimental acceptance.  Radiative and tracking efficiency
corrections have been applied \cite{HBPhD}.  The systematic
uncertainty is 15\%, primarily due to the uncertainty in the low
momentum proton detection efficiency.

The $pp$ and $pn$ pair momentum distributions are similar to each other.  
The $p_{rel}$ distributions rise rapidly starting at $\approx 0.25$ GeV/c
(limited by $p_N \ge 0.25$ GeV/c), peak at $\approx 0.4$  GeV/c,
and have a tail extending to $\approx 0.7$ GeV/c.  The $p_{tot}$
distributions rise rapidly from zero, peak at $\approx 0.25$ GeV/c, and
fall rapidly.  Both distributions have an upper limit determined by
the cut $T_N/\omega \le 0.2$.   These distributions are also
similar for both data sets ($Q^2 \sim 0.7$ \cite{niyazov03} and 1.5 GeV$^2$).
The $Q^2\sim 1.5$ GeV$^2$ $pp$ $p_{rel}$ distribution peaks at slightly larger
momentum  than either the $pn$ or lower $Q^2$ data.  

We also compared our data to a one-body calculation by
Golak, integrated over the experimental acceptance, that
includes an `exact' calculation of the fully correlated initial state
wave function (wf), absorption of the virtual photon by the leading nucleon
and `exact' calculations of the continuum wf of the
spectator $NN$ pair \cite{golak95}.  The calculation does not treat
the rescattering of the leading nucleon.  Including the continuum wf of the $NN$ pair ({\it i.e.,} not treating those two outgoing
nucleons as plane waves) reduces the cross section by about an
order of magnitude.  Note that this calculation is not strictly valid
for $p_{rel} > 0.35$ GeV/c (the pion
production threshold).  This calculation 
significantly underestimates the data.

The one-body calculation of Laget \cite{laget87,laget88,laget04}, using a diagrammatic approach,
sees the same large cross section reduction due to the
$NN$ pair continuum wf.  His one-body calculation
describes the $pn$ pair $p_{rel}$ distribution well. 
Laget's full calculations also indicate large three-body current (MEC
or IC) contributions for both $pn$ and $pp$ pairs.  His three-body
currents improve the agreement for $pp$ pairs and worsen the agreement
for $pn$ pairs.

The ratio of $pp$ to $pn$ spectator pair integrated cross sections is
about 1:4.  This is approximately consistent with the product of the
ratio of the number of pairs and $\sigma_{ep}/\sigma_{en}$, the ratio
of the elementary $ep$ and $en$ cross sections
for $pn$ and $pp$ pairs.
This ratio appears inconsistent with the $pp$ to $pn$ pair
ratio of 1:18 measured in direct pair knockout in $^{12}$C$(e,e'pN)$
\cite{subedi08} at $0.3
< p_{rel} < 0.5$ GeV/c and at relatively low $p_{tot}$ ($<0.15$ GeV/c) .  

\begin{figure}[htbp]
    \includegraphics[height=2.1in]{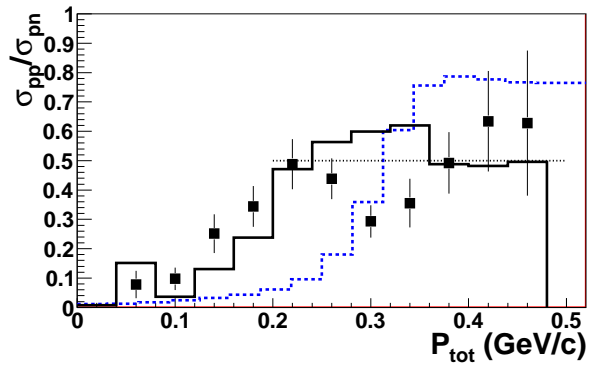}
    \caption{Ratio of $pp$ to $pn$ spectator pair cross sections,
      integrated over $0.3 < p_{rel} < 0.5$ GeV/c.  The points show
      the data, the solid histogram shows the Golak one-body
      calculation \cite{golak95} and the dashed histogram (color online) shows 
      the ratio of the Golak $pp$ and $pn$ bound state momentum distributions.  The dotted line
      at 0.5 shows the simple-minded pair counting result.  The
      data and the one-body calculation have been multiplied by 1.5 to
      approximately account for the ratio of the average
      $ep$ 
      and $en$ elementary cross sections.
      \label{fig:pp_pn_ratio} }
\end{figure}

In order to study this apparent discrepancy we calculated the ratio of
the $pp$ to $pn$ cross sections integrated over $0.3 < p_{rel} < 0.5$
GeV/c as a function of $p_{tot}$ (see Fig.{} 
\ref{fig:pp_pn_ratio}). The ratio has been multiplied by 1.5 to
approximately account for the ratio of the average $ep$ and
$en$ cross sections.  The ratio is very small
for $p_{tot} < 0.1$ GeV/c, consistent with the $^{12}$C$(e,e'pN)$ results,
and increases to 0.4--0.6 for $p_{tot} > 0.2$ GeV/c.  The 
measured ratio increases starting at $p_{tot}\sim 0.1$ GeV/c, in
contrast with that calculated from the bound state wf, which
increases starting at around 0.3 GeV/c.  The ratio is
consistent with  Golak's one-body calculation.  The
ratio  at large $p_{tot}$ is also consistent with simple pair
counting.

This increase in the $pp$ to $pn$ ratio with $p_{tot}$ is a 
signal for the dominance of tensor correlations.  At low $p_{tot}$,
where the angular momentum of the pair with respect to the rest of the
nucleus must be zero, 
the $pp$ pairs predominantly have (isospin,spin) $(T,S)=(1,0)$ \cite{schiavilla07}.  They
are in an $s$-state,
which has a minimum at $p_{rel}\sim
0.4$ GeV/c.  The $pn$ pair is predominantly in a deuteronlike
$(T,S)=(0,1)$ state. Due to the tensor interaction, the $pn$
pair has a significant $d$-state admixture and does not have this minimum
\cite{schiavilla07,sargsian05,alvioli08}. 
This leads to a small ratio at small $p_{tot}$.
As $p_{tot}$ increases, the minimum in the $pp$ $p_{rel}$
distribution fills in, increasing the $pp$ to $pn$ ratio.   


To summarize, we have measured the $^3$He\mbox{$(e,e'pp)n$} reaction
at an incident energy of 4.7 GeV over a wide kinematic range, centered
at $Q^2 \sim 1.5$ GeV$^2$ and $w\approx Q^2/2m_p$.  We selected events
with one leading nucleon and a spectator correlated $NN$ pair by
requiring that the spectator nucleons each have less than 20\% of the
transferred energy and that the leading nucleon's momentum
perpendicular to $\vec q$ be less than 0.3 GeV/c.  The
$p_{rel}$ and $p_{tot}$ distributions for spectator $pp$ and $pn$
pairs are very similar to each other and to those measured at lower
momentum transfer. The ratio of $pp$
to $pn$ pair cross sections for $0.3 < p_{rel} <
0.5$ GeV/c is
very small at low $p_{tot}$ and rises to approximately 0.5 at large
$p_{tot}$.  Since $pp$
pairs at low $p_{tot}$ are in an $s$-state, this ratio
shows the dominance of tensor over central correlations.

\begin{acknowledgments}
We  acknowledge the outstanding efforts of the staff of
the Accelerator and Physics Divisions (especially the CLAS target
group) at Jefferson Lab that made this experiment possible.
This work was supported in part by the Italian Istituto Nazionale di Fisica
Nucleare, the Chilean 
CONICYT, the French Centre National de la Recherche Scientifique and
Commissariat \`{a} l'Energie Atomique, the UK Science and Technology Facilities Council (STFC), the U.S. Department of
Energy and National Science Foundation, 
and the National Research Foundation of Korea.  Jefferson Science Associates, LLC,
 operates the Thomas Jefferson National Accelerator
Facility for the United States Department of Energy under contract
DE-AC05-060R23177.

\end{acknowledgments}


\end{document}